# How does ClinicalTrials.gov Impact Company Innovation?

Yazhou Niu*

*_yazhou.niu@smail.nju.edu.cn_

School of Information Management, Nanjing University, China

**Abstract:** Pharmaceutical companies may have incentives to exaggerate the therapeutic effects of their developed products during the clinical stage, which endangers the health of patients. To increase transparency in clinical practice, the NIH established ClinicalTrials.gov in 2000, which indicates a significant impact on medicine. However, little evidence shows how ClinicalTrials.gov affects medical enterprises' innovation. By identifying the patent application activities through USPTO, Pubmed, and Compustat, we used coherent DID to prove the impact of ClinicalTrials.gov on innovation. We found that the emergence of ClinicalTrials.gov reduced the number of patent applications and led to a shift in R&D directions. This effect can also be moderated depending on firm size, probably because small companies are more incentivized to manipulate data. Hence, we suggest agencies could consider wide-ranging influences when formulating open science policies.

## 1 Introduction

Clinical trial results are golden standards for physicians' clinical decisions (Garattini et al., 2016; Hansson, 2014). Due to the decisive role that clinical trial results play in medical practice, medical companies often publish clinical trial results in academic journals. Through this, medical companies can leverage the academic community to endorse their drugs or medical devices. However, because clinical trial results involve the interests of medical companies, they may have incentives to exaggerate efficacy or conceal side effects (Boutron & Ravaud, 2018; Fanelli, 2009; Li et al., 2018). Therefore, vested interests may lead to bias in clinical trial results

Many scholars have long advocated for increased transparency of clinical trial results (Easterbrook, Gopalan, Berlin, & Matthews, 1991; Simes, 1986). Finally, in 1997, the U.S. Congress passed the FDAMA (Food and Drug Administration Modernization Act), which required the National Institutes of Health (NIH) to create a database of clinical trials to test investigational drugs for serious and life-threatening diseases (Rowberg et al., 1998; Tasneem et al., 2012; Zarin & Tse, 2008; Zarin, Tse, Williams, Califf, & Ide, 2011).

In 2000, NIH officially launched the ClinicalTrials.gov website (NIH, 2000). Then, in 2002 FDA issued a guidance document entitled "Information Program on Clinical Trials for Serious or Life-Threatening Diseases and Conditions." (FDA, 2002) This document guides the industry on procedures for submitting information to the Clinical Trials Data Bank established by section 113 of the Food and Drug Administration Modernization Act of 1997.

These measures had a widespread impact. Research has shown that just the appearance of ClinicalTrials.gov has led to an increase in negative results in studies (Kaplan & Irvin, 2015). This suggests that the emergence of ClinicalTrials.gov has, to some extent, curbed publication bias in scientific research. However, medicine involves not only scientists but also collaboration with the industry, as pharmaceutical or medical companies are largely science-based enterprises (Ahmadpoor & Jones, 2017; McMillan, Narin, & Deeds, 2000). Therefore, a reasonable question arises:

*Does the database of clinical trials influence the industry? And how?*

The previous literature has focused chiefly on knowledge production in scientific laboratories, with little attention to the influence of ClinicalTrials.gov at the enterprise level. This paper utilizes datasets such as USPTO, Pubmed, and Compustat to demonstrate the influence of ClinicalTrials.gov on corporate innovation through coherent DID analysis. Furthermore, considering that companies of different sizes face different reputation constraints, we also discuss the heterogeneity (Guedj & Scharfstein, 2004).

We found that ClinicalTrials.gov significantly influences medical companies' research direction and innovation level. This effect is highly related to a companies' R&D strategies. In addition, this impact is more pronounced in large companies and weaker in small companies, suggesting that ClinicalTrials.gov has a limited regulatory effect on small enterprises. This paper's contribution lies in empirically demonstrating the influence of Clinical Trials on technology and industry, supplementing literature in fields such as innovation, science and technology linkage, open science, etc. We recommend that departments further consider its broad impact when formulating open science policies.

## 2 Data and Methods

### 2.1 Company Finance and Innovation

We first utilized Compustat and the North American Industry Classification System (NAICS) codes to identify companies in the pharmaceutical industry (NAICS code starting with '3254' '541714' '541715') during the period from 1997 to 2005 and obtained relevant financial data of these companies. Drawing from previous literature, we further selected the following variables in Compustat as control variables (Sunder, Sunder, & Zhang, 2017): The logarithm of total assets, Asset-liability ratio, R&D investment, Cash flow, and Retained earnings.

We further utilized patent data from the USPTO to identify medical companies' innovations and patent applications. We extracted Class A and Class C patents from the USPTO based on IPC codes. Using a particular dataset and the Compustat unique identifier for companies, we identified the assignee of these patents. In subsequent analysis, we followed previous studies and measured companies' innovation levels using the number of patent applications rather than granted patents (Dosi, Marengo, & Pasquali, 2006; Seru, 2014). Because the number of patent applications by companies is less likely to be influenced by the patent grant cycle, and patent grants exhibit some instability (Griliches, Pakes, & Hall, 1986). Moreover, patent applications are sufficient to influence companies' performance after the time of application (Ernst, 2001).

We identify the scientific papers cited by these patents. This is because the medical industry is largely science-based, and the patent literature associated with the industry often contains citations to scientific papers (Ahmadpoor & Jones, 2017; McMillan et al., 2000). The characteristics of the scientific literature, to some extent, reflect the technical characteristics of a patent. In our study, for the characteristics of scientific literature cited by patents, we mainly focused on the "basic" or "applied" of scientific literature to examine whether research is in basic medical research or has passed the stage of primary research, and is highly relevant to clinical practice(Ke, 2019). Further, we use the ratio of clinical papers cited in patents filed by the company in the current year as the indicator variable for the direction of innovation.

## 2.2 Method

Medical companies' products, whether drugs or medical devices, require endorsement of their efficacy through clinical trial results. Without sufficient evidence proving the effectiveness of their products, it is challenging for doctors to accept them. Even if a company tries to apply the patenting system to protect its products, the cost of patent protection, such as patent maintenance fees and associated legal expenses, will reduce the cost-effectiveness of protecting products that fail to pass clinical trials. Clinical trials have become more transparent after the introduction of ClinicalTrials.gov and the FDA guidance document, reducing the possibility of companies manipulating clinical trial results. This may affect the results of clinical trials and consequently impact a company's innovation performance. We want to examine whether the appearance of ClinicalTrials.gov and FDA guidance has affected company innovation. Therefore, we have established Model (1),

$$Patent_{it} = \beta_0 + \beta_1 ClincialTrial_{it} + \beta_2 Control_{it} + \gamma_i + \delta_t + \varepsilon_{it} \#(1)$$

where $Patent_{it}$ represents the number of patent applications of firm $i$ in year $t$. $ClincialTrial_{it}$ represents the introduction of FDA guidance. $Control_{it}$ indicates the control variables. $\gamma_i$ and $\delta_t$ represent firm and year fixed effects, respectively.

After the appearance of ClinicalTrials.gov, the direction of enterprise innovation may also be affected due to stricter regulation of clinical research. For example, companies may shift towards basic research to avoid more stringent trial standards. To investigate the effect on research directions, we established Model (2),

$$Per\ Clincial_{it} = \beta_0 + \beta_1 ClincialTrial_{it} + \beta_2 Control_{it} + \gamma_i + \delta_t + \varepsilon_{it} \#(2)$$

where $Per\ Clincial_{it}$ represents the percentage of clinical papers among all papers from PubMed cited by patents files at firm $i$ in year $t$. $ClincialTrial_{it}$ represents the introduction of FDA guidance. $Control_{it}$ indicates the control variables. $\gamma_i$ and $\delta_t$ represent firm and year-fixed effects, respectively.

Since no companies in the United States are unaffected by the advent of ClinicalTrials.gov, and FDAMA only requires registration of clinical results, companies that focus on clinical development rather than basic research are more likely to be affected. Based on this analysis, we design a coherent DID model,

$$Patent_{it} = \beta_0 + \beta_1 \%Clincial_{it} \times I(2001 \leq t) + \beta Control_{it} + \gamma_i + \delta_t + \varepsilon_{it} \#(3)$$

where $\%Clincial_{it} \times I(2001 \leq t)$ represents the interaction term of the percentage of clinical papers among all papers from PubMed cited by patents at firm $i$ in 2001 and the time indicator of the introduction of FDA guidance.

To test the results of Model (3), we further employed a parallel trends test, the

Model is as follows:

$$Patent_{it} = \beta_0 + \sum_{\gamma=1997}^{2005} \beta_{1,\gamma} \%Clincial_{it} \times I(t = \gamma) + \beta_2 Controls_{it} + \gamma_i + \delta_t + \varepsilon_{it} \#(4)$$

Smaller companies are more likely to engage in P-hacking during clinical studies (Guedj & Scharfstein, 2004). Due to their smaller market size and fewer product lines, small companies may face greater business pressure, making them more eager to launch new products. Additionally, being smaller also implies lower visibility, resulting in smaller companies facing fewer reputation constraints (Jin & Leslie, 2009; Mayzlin, Dover, & Chevalier, 2014). Therefore, the manipulation of clinical trial results may be more prevalent among smaller companies. These companies are more likely to be influenced by ClinicalTrials.gov since their trial results are more likely to be unreal. To test whether ClinicalTrials.gov has heterogeneous effects based on company size, we further conducted Model (5),

$$Patent_{it} = \beta_0 + \beta_1 \%Clincial_{it} \times I(2001 \leq t) + \beta_2 \%Clincial_{it} \times I(2001 \leq t) \times Asset_{it}$$
$$+ \beta Control_{it} + \gamma_i + \delta_t + \varepsilon_{it} \#(5)$$

where $\%Clincial_{it} \times I(2001 \leq t) \times Asset_{it}$ indicates the interaction term of the research direction of firm $i$ in 2002, and the year indicates variables of FDA guidance.

# 3 Result

## 3.1 Summary

The medical industry encompasses a wide range of companies, which means enormous differences among them. Therefore, we first report our descriptive statistical results to account for these differences. Table 1 presents the descriptive statistics of the variables.

Table 1. Descriptive statistical analysis.

| VARIABLES | (1) N | (2) Mean | (3) SD | (4) Min | (5) Max |
|---|---|---|---|---|---|
| *Patent* | 2,973 | 14.96 | 52.84 | 1 | 721 |
| *Per Clincial* | 2,739 | 0.712 | 0.251 | 0 | 1 |
| *DA ratio* | 6,950 | 4.432 | 9.553 | 0 | 318.8 |
| *R&D* | 5,726 | 76.96 | 447.8 | -0.0290 | 12,539 |
| *Asset* | 6,925 | 3.164 | 2.395 | -6.908 | 10.76 |
| *Cash* | 6,975 | 107.9 | 592.3 | 0 | 16,055 |
| *Retained income* | 6,697 | -8.492 | 181.5 | -6,763 | 2,593 |

As results show in Table 1, there is considerable variation in patent applications among firms, ranging from 1 to 721, with an average of 14.96. Regarding the research direction of companies, the values of "Per Clinical" range from 0 to 1, with a mean of 0.251. Significant differences were also observed in the other control variables. These indicate substantial variability among companies in our dataset.

## 3.2 The Impact of ClinicalTrials.gov on Corporate Innovation

The sixth column of Table 2 presents the results of Model (1). The regression result shows that ClincialTrial.gov has a negative effect on the number of patent applications ($\beta_1 = -3.305, p < 0.01$). This indicates that after the pre-registration requirement on ClinicalTrials.gov, the average number of patent applications for these companies decreased by 3.305, meaning that ClinicalTrials.gov has had a certain inhibitory effect on companies' innovation.

Table 2. The influence of ClinicalTrials.gov on firms' innovation.

| VARIABLES | (1) $Patent_{it}$ | (2) $Patent_{it}$ | (3) $Patent_{it}$ | (4) $Patent_{it}$ | (5) $Patent_{it}$ |
|---|---|---|---|---|---|
| $ClincialTrial_{it}$ |  | -1.451** | -8.321** | -3.305*** | -3.305*** |
|  |  | (0.02) | (0.02) | (0.00) | (0.00) |
| $DA\ ratio_{it}$ | 0.142 |  | 1.955* | 0.787** | 0.787** |
|  | (0.44) |  | (0.07) | (0.03) | (0.03) |
| $R\&D_{it}$ | 0.00752 |  | -0.0726* | -0.0227*** | -0.0227*** |
|  | (0.34) |  | (0.07) | (0.00) | (0.00) |
| $Asset_{it}$ | 0.0119*** |  | 0.0589*** | 0.0125*** | 0.0125*** |
|  | (0.00) |  | (0.00) | (0.00) | (0.00) |
| $Cash_{it}$ | 0.00127 |  | 0.0111** | 0.00143 | 0.00143 |
|  | (0.20) |  | (0.04) | (0.15) | (0.15) |
| $Retained\ income_{it}$ | -0.0455*** |  | -0.0729*** | -0.0453*** | -0.0453*** |
|  | (0.00) |  | (0.00) | (0.00) | (0.00) |
| Constant | 12.52*** | 16.17*** | 1.335 | 11.58*** | 11.58*** |
|  | (0.00) | (0.00) | (0.62) | (0.00) | (0.00) |
| Observations | 2,505 | 2,875 | 2,594 | 2,505 | 2,505 |
| R-squared | 0.927 | 0.897 | 0.686 | 0.927 | 0.927 |
| Firm FE | YES | YES | NO | YES | YES |
| Industry FE | YES | YES | YES | NO | YES |

Standard errors clustered by industries (NAICS) in parentheses.
*** p < 0.01, ** p < 0.05, * p < 0.1.

After the FDA guidance has been approved, companies should comply with the requirements of ClinicalTrials.gov. Therefore, companies need to adjust their previous research pipelines to comply with the new regulation procedure, leading to a decrease in the number of patent applications. In addition, ClinicalTrials.gov reduces companies' manipulation space for results, decreasing the "efficiency" of R&D. Another possible reason is that ClinicalTrials.gov avoids overlapping research pipelines among companies by publicly disclosing their research information and progress. This, in turn, may influence patent applications.

## 3.3 The Impact of ClinicalTrials.gov on Corporate Research Direction

Will the ClinicalTrials.gov affect the innovation direction of enterprises? Table 3 demonstrates the result of Model (2) and the answer. The regression result shows that ClincialTrial.gov has a negative effect on the R&D direction ($\beta_1 =- 0.0344, p < 0.05$).This indicates that after the appearance of ClinicalTrials.gov, the percentage of clinical papers cited in patents applied by these companies decreased by 3.4%.

Table 3. The influence of ClinicalTrials.gov on firms' innovation direction.

| VARIABLES | (1) Per Clincial$_{it}$ | (2) Per Clincial$_{it}$ | (3) Per Clincial$_{it}$ | (4) Per Clincial$_{it}$ | (5) Per Clincial$_{it}$ |
|---|---|---|---|---|---|
| ClincialTrial$_{it}$ |  | -0.0373*** | -0.0771*** | -0.0344** | -0.0344** |
|  |  | (0.00) | (0.00) | (0.01) | (0.01) |
| DA ratio$_{it}$ | -0.0144*** |  | 0.0127** | -0.00736** | -0.00736** |
|  | (0.01) |  | (0.01) | (0.01) | (0.01) |
| R&D$_{it}$ | 0.00137 |  | -0.00192*** | 0.000847 | 0.000847 |
|  | (0.21) |  | (0.00) | (0.46) | (0.46) |
| Asset$_{it}$ | 1.74e-06 |  | 1.98e-05*** | 7.64e-06* | 7.64e-06* |
|  | (0.74) |  | (0.00) | (0.08) | (0.08) |
| Cash$_{it}$ | -6.16e-06*** |  | -9.23e-06** | -4.67e-06** | -4.67e-06** |
|  | (0.01) |  | (0.02) | (0.02) | (0.02) |
| Retained income$_{it}$ | -6.45e-06** |  | -3.60e-05*** | -4.12e-06* | -4.12e-06* |
|  | (0.02) |  | (0.00) | (0.09) | (0.09) |
| Constant | 0.772*** | 0.736*** | 0.706*** | 0.761*** | 0.761*** |
|  | (0.00) | (0.00) | (0.00) | (0.00) | (0.00) |
| Observations | 2,294 | 2,641 | 2,388 | 2,294 | 2,294 |
| R-squared | 0.581 | 0.597 | 0.089 | 0.584 | 0.584 |
| Firm FE | YES | YES | NO | YES | YES |
| Industry FE | YES | YES | YES | NO | YES |

Standard errors clustered by industries (NAICS) in parentheses.
*** $p < 0.01$, ** $p < 0.05$, * $p < 0.1$.

This suggests that ClinicalTrials.gov affects the output of companies' innovation and influences the direction of R&D. This result is not surprising. When clinical research is limited, they naturally tend to or have to focus on basic research instead.

## 3.4 The Casual Effect of ClinicalTrials.gov on Corporate Innovation

We further designed a coherent DID model based on the innovation directions of companies to demonstrate the causal effect of ClinicalTrials.gov, thus eliminating potential confounding factors. Table 4 displays the regression results.

Table 4. The influence of ClinicalTrials.gov on firms' innovation based on different R&D strategies.

| VARIABLES | (1) $Patent_{it}$ | (2) $Patent_{it}$ | (3) $Patent_{it}$ | (4) $Patent_{it}$ | (5) $Patent_{it}$ |
|---|---|---|---|---|---|
| $\%Clincial_{it} \times I(2001 \leq t)$ | -0.941*** | -0.854*** | 0.197*** | -0.373*** | -0.642*** |
|  | (0.00) | (0.00) | (0.00) | (0.00) | (0.00) |
| $DA\ ratio_{it}$ |  | 0.531*** | 0.515*** | 0.152*** | 0.192*** |
|  |  | (0.00) | (0.00) | (0.00) | (0.00) |
| $R\&D_{it}$ |  | -0.0350*** | -0.0345*** | -0.00728** | -0.0119** |
|  |  | (0.00) | (0.00) | (0.03) | (0.01) |
| $Asset_{it}$ |  | 0.000165*** | 0.000178*** | 7.64e-05*** | 7.93e-05*** |
|  |  | (0.00) | (0.00) | (0.00) | (0.00) |
| $Cash_{it}$ |  | 3.10e-05*** | 6.67e-05*** | -4.00e-07 | 1.89e-05*** |
|  |  | (0.00) | (0.00) | (0.90) | (0.00) |
| $Retained\ income_{it}$ |  | -0.000269*** | -0.000240*** | -0.000192*** | -0.000133*** |
|  |  | (0.00) | (0.00) | (0.00) | (0.00) |
| Constant | 4.789*** | 0.134 | -0.200*** | 3.282*** | 3.054*** |
|  | (0.00) | (0.14) | (0.00) | (0.00) | (0.00) |
| Observations | 1,958 | 1,733 | 1,733 | 1,726 | 1,726 |
| Firm FE | YES | NO | NO | YES | YES |
| Year FE | YES | NO | YES | NO | YES |

Standard errors clustered by industries (NAICS) in parentheses.
*** p < 0.01, ** p < 0.05, * p < 0.1.

According to the regression results in the sixth column, the coefficient of policy intensity shows an inhibition effect on companies' innovation levels ($\beta_1 = -0.642, p < 0.01$). This indicates that the decrease in patent applications by medical companies after the introduction of ClinicalTrials.gov and FDA guidance is closely related to the company's research direction. Specifically, the more a company's research tended towards clinical directions before the appearance of FDA guidance, the greater the impact it experienced from ClinicalTrials.gov, leading to a greater effect on its patent application.

## 3.5 Robustness Check

Figure 1 shows the results of the parallel trends test. It is observed that before the policy implementation, the impact of companies' research direction in 2001 on their patent application volume was not significant. However, after the policy implementation, these effects present a sharp decline.

Figure 1: Results of parallel trend test.

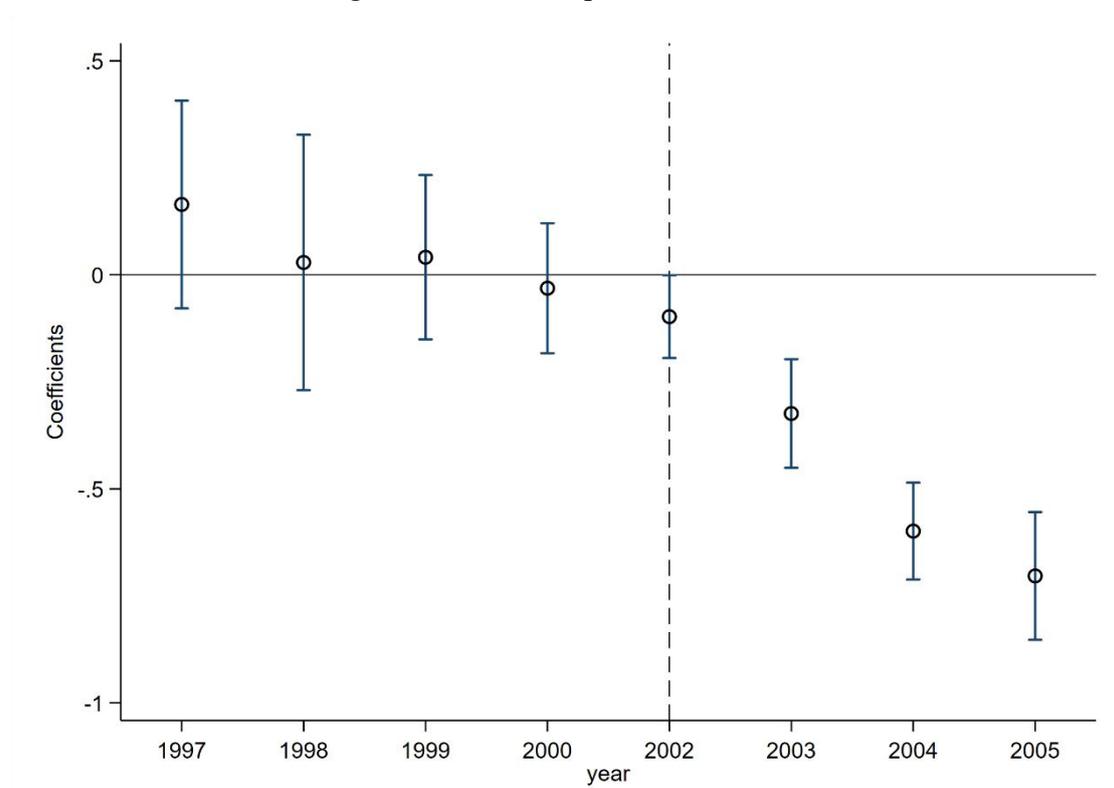

Our result indicates that the influence of ClinicalTrials.gov is unlikely to be related to unobserved events that may coincide with ClinicalTrials.gov, and the results of Model (3) are robust.

## 3.6 The Moderation Effect of ClinicalTrials.gov Depending on Firm Size

Table 5. Heterogeneity of ClinicalTrials.gov at firm size level.

| VARIABLES | (1) $Patent_{it}$ | (2) $Patent_{it}$ | (3) $Patent_{it}$ |
|---|---|---|---|
| $\%Clincial_{it} \times I(2001 \leq t)$ |  | -0.642*** | -0.931*** |
|  |  | (0.00) | (0.00) |
| $Asset_{it}$ | 0.169*** | 0.192*** | 0.189*** |
|  | (0.00) | (0.00) | (0.00) |
| $\%Clincial_{it} \times I(2001 \leq t) \times Asset_{it}$ |  |  | 0.0524** |
|  |  |  | (0.03) |
| $DA\ ratio_{it}$ | -0.00753*** | -0.0119** | -0.0147*** |
|  | (0.01) | (0.01) | (0.01) |
| $R\&D_{it}$ | 8.55e-05*** | 7.93e-05*** | 6.45e-05*** |
|  | (0.00) | (0.00) | (0.00) |
| $Cash_{it}$ | 2.12e-05*** | 1.89e-05*** | 1.35e-05*** |
|  | (0.00) | (0.00) | (0.00) |
| $Retained\ income_{it}$ | -0.000142*** | -0.000133*** | -0.000139*** |
|  | (0.00) | (0.00) | (0.00) |
| Constant | 2.830*** | 3.054*** | 3.068*** |
|  | (0.00) | (0.00) | (0.00) |
|  |  |  |  |
| Observations | 2,505 | 1,726 | 1,726 |
| Firm FE | YES | YES | YES |
| Year FE | YES | YES | YES |

Standard errors clustered by industries (NAICS) in parentheses.
*** p < 0.01, ** p < 0.05, * p < 0.1.

 The fourth column of Table 5 shows the regression results for Model (5). The coefficient of the interaction term between policy intensity and companies' assets positively affects the number of patent applications ($\beta_1 = 0.0524, p < 0.05$). Its coefficient is opposite in direction to $\beta_1$, indicating that company size has an inhibitory effect on the impact of ClinicalTrials.gov. This result is consistent with the previous analysis; smaller companies are more likely to engage in result manipulation. Thus, ClinicalTrials.gov is more likely to suppress their patent activities.

## 4 Conclusion

ClinicalTrials.gov plays a significant role in the development of the medical industry. While previous studies have largely focused on how ClinicalTrials.gov influences scientific production, our research suggests that it also profoundly impacts companies' innovation.

After the emergence of ClinicalTrials.gov and the requirement of pre-registration by the FDA, medical companies' innovations were suppressed. Moreover, this suppression was highly correlated with the companies' previous R&D strategies. In addition, this suppression manifested heterogeneity at the firm level, with smaller companies experiencing a more substantial inhibitory effect, perhaps because smaller companies have stronger incentives and lower manipulation costs. Therefore, the inhibition of enterprise innovation by ClinicalTrials.gov may not imply a negative impact. Instead, it is more likely to suppress companies' p-value manipulation.

This study also has limitations. We did not identify the impact of ClinicalTrials.gov on patent value, which may be evaluated in the future based on other bibliometric indicators. Additionally, we did not explore the effects of ClinicalTrials.gov on other companies' financial metrics, such as profitability and market value, which may be addressed in future research.

Based on our analysis, we recommend that departments, when formulating open science policies, should consider their wide-ranging impact. They should not solely focus on science but also consider the effects on technology and industry, especially for those industries based on science. Additionally, oversight of small companies should be strengthened, as they have a stronger motivation to manipulate data.

## Open science practices

Data and codes can be obtained by email, please indicate your intention when sending.

## Competing interests

No competing interests.